\documentstyle[aps,pra]{revtex}
\begin{document}
\twocolumn[
\begin{centering}
{\Large \bf Formation of antihydrogen atoms in an ultra-cold
positron-antiproton plasma}\vspace{1.5ex}\\
P. O. Fedichev\\
\vspace{1ex}
\begin{center}
{\it FOM Institute for Atomic and Molecular Physics, Kruislaan 407,
 1098 SJ Amsterdam, The Netherlands}
\end{center}
\end{centering}
\begin{quote}
We discuss the formation of antihydrogen atoms ($\bar{\rm H}$)
in an ultra-cold positron-antiproton plasma.
For positron densities $n_p\agt 10^8$ cm$^{-3}$ the
characteristic formation time of stable $\bar{\rm H}$ is determined
by collisional relaxation of highly excited atoms produced in the
process of 3-body Thompson recombination.
Relying on the mechanisms of ``replacement collisions'' and ``
transverse collisional drift'' we find a bottleneck in the relaxation
kinetics and analyze the physical consequences of this phenomenon.
\end{quote}]
\vspace{4mm}

\narrowtext
Recently, the bound state of an antiproton with a positron, i.e.
antihydrogen
($\bar{\rm H}$), has been produced at relativistic energies
\cite{Oelert}.
Succesful experiments on accumulating up to $10^5$ antiprotons
in an ion trap at temperatures $T\sim 4K$ \cite{Jerry1}
initiated the ideas to create and study ultra-cold $\bar{\rm H}$.
As discussed in \cite{Hughes,HZ,Jerry2}, antihydrogen at sub-Kelvin
energies is an ideal system for testing
CPT-invariance and the weak equivalence principle
by gravity and spectroscopic experiments.

Cold antihydrogen is expected to be produced by merging cold plasmas of
positrons and antiprotons \cite{Jerry3}.
The attention is now mainly focused on the formation of $\bar{\rm H}$
through the process of Thompson capture in 3-body collisions
\begin{equation}
\label{TC}
\bar{p}+e^++e^+\rightarrow \bar{\rm H}^*+e^+.
\end{equation}
Neutral (anti)atoms $\bar{\rm H}^*$ produced in this process have
binding
energies $\varepsilon\sim T$ and a characteristic size of the order of
the Thompson radius $R_T=e^2/T$ ($e$ is the positron charge).
They subsequently undergo collisional
and radiative relaxation and are supposed to acquire binding energies
larger than the threshold energy of ionization by the electric field at
the ends
of the ion trap.
For realistic fields ($10-100$ V/cm) the threshold energy
$\varepsilon_0$ varies from $20$ to $60$K,
which corresponds to the principal quantum number $n_0\sim 100-50$.
We will use the term ``stable'' for atoms with binding energies
$\varepsilon>\varepsilon_0$, as they leave the ion trap without
ionization and can be further captured in a neutral atom trap.

We will discuss the case of an ideal plasma where $T\gg e^2n_e^{1/3}$
and, hence, we have a small parameter
\begin{equation}           \label{smpar}
n_eR_T^3\ll 1.
\end{equation}
Then the event rate of the Thompson process, $\alpha
n_pn_e^2$ ($n_p$ and $n_e$ are the densities of antiprotons and
positrons), is determined by the rate constant
\begin{equation}
\label{TCrate}
\alpha=A\frac{e^{10}}{\sqrt{m}T^{9/2}},
\end{equation}
with $m$ being the positron mass. The numerical coefficient $A$ ranges
from $0.8$ in zero magnetic field (see, e.g., \cite{Keck}) to $0.08$ in 
infinitely high fields \cite{O'Neil}.
The rate rapidly increases with decreasing $T$ and
for sufficiently high positron densities $n\sim 10^8$ cm$^{-3}$
already at temperatures of a few Kelvin the Thompson
process becomes very fast. However, the formation of stable
($\varepsilon>\varepsilon_0$) ultra-cold $\bar{\rm H}$ is governed by
the relaxation kinetics. Thus, irrespective of the ratio between the
corresponding relaxation time $\tau$ and the characteristic time of
Thompson capture $\tau_T\sim 1/\alpha n_e^2$, the former is the key
parameter of the problem.

For high magnetic fields ($B\agt 5$ T) and low temperatures ($T\sim 4$
K) used in antiproton experiments \cite{Jerry1,Jerry2,Jerry3},
the size $r$ of atoms with binding energies
$\varepsilon\alt\varepsilon_0$,
formed in the Thompson process and undergoing relaxation,
satisfies the inequality
\begin{equation}            \label{TB}
R_T\agt r\gg r_B,
\end{equation}
where $r_B=v_T/\omega_B$ is the positron Larmor radius, $v_T=\sqrt{T/m}$ 
the thermal velocity, and $\omega_B=eB/mc$ the Larmor frequency.
Under the condition (\ref{TB})
the relaxation is strongly influenced by the field.
At experimental positron densities the formation of stable
$\bar{\rm H}$ by collisional relaxation dominates over radiative
deexcitation.
Considering both the mechanisms of ``replacement collisions''
\cite{O'Neil}
and ``transverse collisional drift'' \cite{FM95} we find a bottleneck
in the relaxation kinetics and discuss the physical consequencies of
this phenomenon.

As we consider the relaxation of highly excited
neutral atoms with principal quantum numbers $n\agt n_0\gg 1$,
the system can be treated classically.
The criterion (\ref{TB}) allows us to use the drift approximation
(see e.g. \cite{LLX}) for the motion of a bound positron:
In the plane transverse to the magnetic field the positron moves in
a circle of radius $r_B$, and the center of the circle (oscillating
along the magnetic field direction $z$ in the Coulomb potential well
$U(\rho,z)=-e^2/\sqrt{z^2+\rho^2}$~)
${\bf E}\times {\bf B}$ drifts around the antiproton at a fixed
transverse distance $\rho$.
This approximation assumes that the drift frequency
$\omega_d\sim ce/B\rho^3\ll \omega_B$ and, hence,
\begin{equation}\label{regchaosborder}
\rho\gg R_B\equiv\left(\frac{mc^2}{B^2}\right)^{1/3}.
\end{equation}
The condition (\ref{TB}) ensures that $r_B\ll R_B\ll R_T$.
In high magnetic fields the distance $R_B$ is significantly
smaller than the characteristic size of the atom with the binding
energy $\varepsilon\approx\varepsilon_0$ and, hence, the formation of
stable
$\bar{\rm H}$ can be well described by using the drift approximation for
the positron motion.

Relaxation of highly excited atoms with the transverse size $\rho$
satisfying the condition (\ref{regchaosborder}) proceeds through
collisions with plasma positrons and involves two leading mechanisms.
The mechanism of ``replacement collisions'', first found in numerical
calculation \cite{O'Neil}, assumes that an incident plasma positron
has an impact parameter smaller than $\rho$ and becomes bound with the
antiproton in the course of the collision, whereas the initially bound
positron becomes unbound and carries away the increase of the binding
energy.
The latter is large ($\sim e^2/\rho$) and, hence, replacement
collisions provide a fast relaxation of the positron longitudinal motion
in the Coloumb well $U(\rho,z)$.
Already after a few replacement collisions the
longitudinal motion comes to thermal equilibrium with the background
plasma
and the positron oscillates in the bottom of the Coulomb well, with
the characteristic amplitude $z_T\sim(\rho^3/R_T)^{1/2}\ll \rho$.
Hence, the atom size $r\approx\rho$,
and the binding energy is related to the transverse size
by $\varepsilon=e^2/\rho$.

The most important consequence of replacement collisions is the
increase of the binding energy of the atom and, accordingly, decrease
of the atom size $\rho$.
The frequency of replacement collisions is $\sim n_e \rho^2 v_T$ and 
the change of the size in each collision is of order $\rho$ itself. 
Therefore, the time dependence of the size can be found from the
equation
\begin{equation} \label{replv}
\dot\rho^{(r)}\approx -n_ev_T\rho^3.
\end{equation}
For $\dot\rho$, $\rho$ treated as ensemble averages the validity
of Eq.(\ref{replv}) at times $t\gg \tau_R=1/n_ev_TR_T^2$ can be 
proven by direct solution of the kinetic equation.

The rate of replacement collisions is independent of the magnetic
field.
These collisions provide efficient relaxation of excited atoms
with large $\rho$ (small binding energy) even in infinitely high fields.
At finite $B$ and sufficiently small $\rho$, but still satisfying the
condition (\ref{regchaosborder}), there is another relaxation mechanism,
``transverse collisional drift'' resulting from far
positron-atom collisions with impact parameters larger than $R_T$
\cite{FM95}.
Due to such collisions a bound positron undergoes diffusion in the plane
perpendicular to the magnetic field. The corresponding diffusion
coefficient is the same as that for free plasma positrons due to far
positron-positron collisions:
\begin{equation}     \label{Dperp}
D_{\perp}=\sqrt{\frac{\pi}{2}}\Lambda^2\frac{r_B^2}{\tau_R},
\end{equation}
where $\Lambda=\log{(n_eR_T^3)^{-1/3}}$ is the Coulomb logarithm.
The principal difference originates from the fact
that the
longitudinal motion of bound positrons is coupled to the transverse one,
whereas for free positrons the coupling is absent. Therefore,
collisions between free positrons do not change their longitudinal
velocities and, hence, do not contribute to the mobility \cite{LLX}.
On the contrary, far
positron-atom collisions change both the longitudinal velocity of a
bound
positron and its distance from the antiproton. This leads to a non-zero
mobility coefficient of bound positrons, $b_{\perp}=D_{\perp}/T$
\cite{FM95}.
Thus, there will be a collision-induced drift of the bound positron
towards the antiproton, with the velocity
\begin{equation} \label{coldriftv}
\dot\rho^{(d)}=-b_{\perp}\frac{e^2}{\rho^2}.
\end{equation}
The drift increases the binding energy of excited atoms and, in
contrast to replacement collisions, provides efficient relaxation at
small $\rho$.

On the basis of the two described mechanisms, we can now draw the
picture of the kinetics of collisional relaxation. For $\rho\sim R_T$
the ratio of the rate of replacement collisions to the rate of
transverse collisional drift is $\sim(R_T/r_B)^2$. Hence, highly
excited atoms with binding energies $\sim T$, formed in the Thompson
process (\ref{TC}), first undergo relaxation due to replacement
collisions. As follows from Eqs.~(\ref{replv}) and
(\ref{coldriftv}), the rate of replacement collisions decreases 
with increasing binding energy $\varepsilon$ (decreasing $\rho$) and 
at $\rho\approx\rho_b=R_B(R_T/R_B)^{2/5}\ll R_T$
becomes equal to the rate of transverse collisional drift. Further
increase of $\varepsilon$ occurs due to the transverse drift,
and the relaxation rate increases with $\varepsilon$. Thus, the total
relaxation rate is minimum at $\rho\approx\rho_b$, and this point
actually represents a bottleneck in the relaxation kinetics.
The ratio $\rho_b/R_B\gg 1$ justifies the use of the
drift approximation for $\rho\sim \rho_b$.
The characteristic time of the formation of atoms with size $\rho>R_B$
is given by $\tau=\int_{\rho}^{R_T}d\rho/\dot\rho$, where
$\dot\rho=\dot\rho^{(r)}+\dot\rho^{(d)}$.
The main contribution to the integral comes from distances
$\rho\sim \rho_b$, and $\tau$ becomes independent of $\rho$:
\begin{equation}           \label{taurelB}
\tau\approx\left(\frac{R_T}{r_B}\right)^{4/5}\tau_R\propto
T^{3/10}B^{4/5}.
\end{equation}

In the range of fields and
temperatures relevant for producing antihydrogen in ion traps
\cite{Jerry3}), the bottleneck binding energy 
$\varepsilon_b=e^2/\rho_b\propto (BT)^{2/5}$ is smaller or
of the order of the threshold energy $\varepsilon_0$ for typical 
experimental conditions.
Therefore, the formation of stable $\bar{\rm H}$ will be determined by
the characteristic time $\tau$ (\ref{taurelB}). It is important to
emphasize that possible decrease of temperature or change of the
magnetic field in ion trap experiments can not appreciably reduce
$\tau$. One can also check that the characteristic time of radiative
relaxation of atoms with binding energies
$\varepsilon\alt\varepsilon_0$ is much larger than $\tau$.

Neutral atoms are produced in the inner spatial region of the ion trap,
where the electric field is screened by the plasma, and move freely
with velocities $v_a\sim\sqrt{T/M}$ ( $M$ is the atom mass).
If the size $L$ of the inner region is such that the transit time
$\tau_L\ll \tau$, then the binding energy of atoms which reach the 
ends of the trap is smaller than $\varepsilon_0$. The atoms undergo 
ionization by the electric field
and return to the plasma as positrons and antiprotons. In this case
the outcome of antihydrogen from the trap is negligible.
Promising is the opposite case $\tau_L\gg \tau$, where the atoms reach
the ends of the trap after acquiring binding energies
$\varepsilon>\varepsilon_0$ and escape from the trap without ionization.
Then the rate of producing neutral $\bar{\rm H}$ by the ion trap is
determined by the rate of Thompson capture $N_p/\tau_T$ ($N_p$ is the
number of antiprotons).
For realistic parameters of the ion trap experiments \cite{Jerry3}:
$n_e=10^8$ cm$^{-3}$, $T=4$ K, and $B=10$ T, we have $\tau_T\approx
10^{-4}$ s, $\tau\approx 10^{-5}$ s, and $v_a\approx 2\times 10^4$ cm/s.
The condition $\tau_L/\tau\gg 1$ requires $L\sim 1$ cm.
As the ratio $\tau_L/\tau\propto T^{-4/5}$, and $\tau_T\propto T^{-9/2}$
the decrease of temperature to
$1$ K allows to have $L$ of several millimeters and reduces $\tau_T$ by
more
than $2$ orders of magnitude.

The author is grateful to G. Gabrielse for stimulating the interest to
the 
subject of this paper and helpful discussions.
Fruitful discussions with T. M. O'Neil, G.V. Shlyapnikov, J.T.M.
Walraven
and A. Mosk are acknowledged. This work was supported by the Stichting
voor 
Fundamenteel Onderzoek der Materie (FOM) and by the Grant from the
Smitsonian
Astrophysical Observatory's Institute for Theoretical Atomic and
Molecular
Physics.

\end{document}